\documentclass[twocolumn,amsmath,amssymb,aps,pre,10pt,showpacs]{revtex4-1}

\usepackage{graphicx}

\usepackage{amsmath}

\usepackage[dvips]{color}

\begin{document}

\title{A metric approach for sound propagation in nematic liquid crystals}

\author{E. Pereira}
\email{Corresponding author: erms@fis.ufal.br}
\affiliation{Instituto de F\'{i}sica, Universidade Federal de Alagoas, Campus A.C. Sim\~{o}es, 57072-900,
Macei\'{o}, AL, Brazil.}
\author{S. Fumeron}
\affiliation{Laboratoire d'\'Energ\'etique et de M\'ecanique Th\'eorique et Appliqu\'ee, 
CNRS UMR 7563, Nancy Universit\'e, 54506, Vand\oe{uvre} Cedex, France.}
\author{F. Moraes}
\affiliation{Departamento de F\'{\i}sica, CCEN,  Universidade Federal da Para\'{\i}ba, 58051-970, Caixa Postal
5008, Jo\~ao Pessoa, PB, Brazil.}
\begin{abstract}

In the eikonal approach, we describe sound propagation near to topological defects of nematic liquid crystal as geodesics of a non-euclidian manifold endowed
with an effective metric tensor. The relation between the acoustics of the medium and this geometrical
description is given by Fermat's principle. We calculate the ray trajectories and propose a diffraction experiment to retrieve informations about the elastic constants.
\end{abstract}


\maketitle

\section{Introduction}
Number of systems in condensed matter physics can be described by the same mathematical structures as those describing 
the gravitational field in General Relativity: this is the core of the analogue gravity (or effective geometry)
programme \cite{visser}. In General Relativity, gravitation is accounted by the distortions of spacetime, which is
modeled by a pseudo-riemannian \textit{manifold} \cite{weinberg} of signature (-,+,+,+). The properties of this
spacetime are encompassed in the metric tensor \textbf{g} (or simply \textit{metric}) and other quantities derived from
it, such as the Riemann tensor or the Ricci scalar. Free-falling particles follow trajectories corresponding to the
geodesics (curves of extremal length) of the manifold. In particular, light paths correspond to the null geodesics of
spacetime
\begin{align}
ds^2=g_{\mu\nu}dx^\mu dx^\nu=0,
\end{align}
where $ds^2$ is the square of elementary length along the geodesic and the $g_{\mu\nu}$ are the covariant components of
\textbf{g} (the Einstein summation convention is assumed, where greek indices refer to spacetime
coordinates and latin indices are related to spatial ones).

As light trajectories {can also be curved when crossing a refractive medium}, Fermat's principle can be stated in terms of an effective geometry \cite{weinberg,alsing}. For a given refractive index $n(\vec{r})$, where
$\vec{r}$ is
the position vector of the light wavefront, the light trajectories are interpreted as the geodesics of a non-Euclidean
space with the line element d$\Sigma^2$
\begin{equation}\label{2}
 d\Sigma^2=n^2(\vec{r}) (dx^2+dy^2+dz^2),
\end{equation} 
Despite (\ref{2}) represents only a three-dimensional non-Euclidean manifold, the corresponding geodesics are equal to the null ones in a static four-dimensional pseudo-riemannian manifold with the time coordinate \cite{misner}. Since we will deal with static refractive indexes, whenever considering the time, the time coordinate appears in the line element 
\begin{align}\label{time}
	ds^2=-c^2dt^2+d\Sigma^2,
\end{align}
where $c$ is the speed of the light in vacuum  and $d\Sigma^2$
represents the spatial line element obtained by the method that will be exposed in this paper.  For
the study that will be developed here, $c$ is the velocity of the sound in the material of interest when it is
homogeneous and isotropic.

In this paper, we will apply this gravitational analogy to describe the sound propagation in liquid crystals \cite{stephen,gennes,kleman}, materials raising a considerable interest as they are both theoretically challenging \cite{bowick,erms2,kapustina} and of practical interest (PVA screens, q-plates or cellphones \cite{miyata}). They are made of anisotropic molecules (disc-like or rod-like) that present meta-states
between the crystalline and liquid phases. One of these meta-states is the nematic phase: the molecules' axis are
oriented on average along a specific direction called \textit{director} (represented by the versor $\hat{n}$) whereas their centers of mass are randomly located. Once in the nematic phase, the refractive index, the sound velocity and other macroscopic properties become anisotropic. Moreover, the nematic phase can exhibit birefringence, as the director plays the role of the optical axis \cite{mason,stephen, gennes}. 

The gravitational analogy will be based on Fermat's principle, once it can also be used to determine sound trajectories \cite{horz}. The local speed of sound plays the role of the speed of light in General Relativity and sound propagates along the null geodesics of the acoustic metric. The {interest of using this analogy (or, more properly, this geometrical approach) is that it helps understanding from a different point of view several problems dealing} with acoustics (for example, tomographic image reconstruction via ray tracing \cite{wang}, flight time in acoustical waveguides \cite{virovlyansky}, problems using variational method \cite{mi}) or nematoacustics \cite{virga}. 

We will consider the sound propagation around a punctual (called \textit{hedghog}) and a linear 
(called \textit{disclination}) topological defect of the nematic phase of a liquid crystal with rod-like molecules \cite{gennes,kleman, stephen,kleman2}. Sound trajectories will
be exhibited and the diffraction patterns will be determined in the perspective of retrieving the elastic constants of
the liquid crystalline medium.

\section{Analog model of anisotropic liquid crystals}

\subsection{Acoustic aspects}

The specific structure of the liquid crystal in nematic phase allows two kinds of acoustic waves to exist \cite{mullen,stephen,gennes}, similarly for sound in {solid} crystals \cite{fedorov,royer}: the ordinary wave, that behaves as inside an isotropic medium, and  the \emph{extraordinary} one that depends on the angle between the direction of the propagation and the director $\hat{n}$. For each kind of wave, one usually defines two characteristic velocities: the phase velocity, which points in the direction of the wave vector $\vec{k}$, and the \emph{group velocity} that is oriented similarly to the acoustic Poynting vector $\vec{S}$ (in general, not parallel to $\vec{k}$) and indicates the direction of energy propagation. The phase
velocity $v_p$ of the extraordinary wave, for a liquid crystal in the nematic phase with the director $\hat{n}$ orientated on the z-direction, can be measured by the pulse-superposition technique \cite{mason}, producing the expression 
\begin{align}
\label{vp}
	v_p^2(\alpha)=\frac{\left(C_{11}+\left(C_{33}-C_{11}\right)\cos^2\alpha\right)}{\rho},
\end{align}
where $\rho$ is the density, $\alpha$ is the angle between the wave vector $\vec{k}$ and $\hat{n}$ \cite{stephen,gennes}. {This relation expresses that a liquid crystal can be viewed }as a simple solid with linear elastic constants in the x- and y- directions both equal to $C_{11}$ and in the z-direction equal to $C_{33}$ \cite{mullen,stephen}. Additionally, {the smallest portion of liquid crystal where the director can be defined presents local cylindrical symmetry and therefore, we will consider (\ref{vp}) to be locally valid for any configuration of the director $\hat{n}$ (even in the presence of defects as in sections \ref{app1} and \ref{app2})}. 


The procedure to obtain $N_g(\beta)$ relies firstly on the {determination} of $v_p(\alpha)$ and phase refractive index $N_p(\alpha)$. To determine the latter, one replaces (\ref{vp}) in $N_p^2=\frac{v^2}{v_p^2(\alpha)}$, where $v$ is the velocity of sound in the
isotropic phase of the liquid crystal, obtaining
\begin{align*}
 N_p^2=\frac{v^2\rho}{C_{11}\sin^2\alpha+C_{33}\cos^2\alpha},
\end{align*} 
or
\begin{align}
	\frac{N_p^2\sin^2\alpha}{\frac{v\rho}{C_{11}}}+\frac{N_p^2\cos^2\alpha}{\frac{v\rho}{C_{33}}}=v. \label{zut}
\end{align}
Following \cite{kleman}, we define the refractive index vector $\vec{N_p}\equiv\frac{v\hat{k}}{v_p}$ and denote by index $_\bot$ (resp. $_{//}$) components of vectors that are orthogonal (resp. parallel) to director $\hat{n}$. Thus, the components of $\vec{N_p}$ are $N_{p_\bot}=N_p \sin\alpha$ and $N_{p_{//}}=N_p \cos\alpha$. For acoustic waves in anisotropic crystals (which is the case of a liquid crystal in its nematic phase \cite{born,fedorov}), it is known that
\begin{align}
 \vec{v_g}\cdot\hat{k}=v_p\Longrightarrow\vec{v_g}\cdot\vec{N_p}=v.\label{con}
\end{align}
Identifying (\ref{con}) with (\ref{zut}), we found the components of $\vec{v_g}$
\begin{align}
\label{consi}
	 v_{g_\bot}=\frac{N_{p_\bot}}{\frac{v\rho}{C_{11}}}; \;\;\; v_{g_{//}}=\frac{N_{p_{//}}}{\frac{v\rho}{C_{33}}}.
\end{align}
Introducing $\beta$ as the angle between $\vec{v_g}$ and $\hat{n}$, then alternate expressions for the group velocity components are $v_{g_\bot}=v/N_g \sin\beta$ and $v_{g_{//}}=v/N_g \cos\beta$. Thus, the use of (\ref{consi}) in (\ref{con}) results in
\begin{eqnarray}
v&=&\vec{v_g}\cdot\vec{N_p}=\frac{v}{N_g}\left(N_{p_\bot}\sin\beta+N_{p_{//}}\cos\beta\right) \nonumber \\
&=&\frac{v^2}{N_g^2}\left(\frac{v\rho}{C_{11}}\sin^2\beta+\frac{v\rho}{C_{33}}\cos^2\beta\right). \nonumber
\end{eqnarray}
Finally, we obtain
\begin{align}
\label{extra}	N_g^2(\beta)=\frac{v^2\rho}{C_{11}}\sin^2\beta+\frac{v^2\rho}{C_{33}}\cos^2\beta.
\end{align}

In the next section, we will discuss about the connection between the acoustic group index and the effective metric experienced by sound in the nematics.



\subsection{Geometric aspects}

We intent to study the paths followed by the extraordinary ray. This can be achieved by applying Fermat's principle \cite{Kline}, for which the path followed by sound between points $A$ and $B$ is the one that minimizes the integral
\begin{align}\label{eq1}
	F=\int_A^B N_g(\beta)dl,
\end{align}
where $l$ is the arc length of the path. As pointed out by Joets and Ribotta \cite{Joets} for light inside an anisotropic crystal, the use of Fermat's principle to determine sound paths is equivalent to calculating the null geodesics of a manifold with line element $d\Sigma^2=\sum_{i,j}{g_{ij}dx^idx^j}$, where this null geodesic is the curve that minimizes the integral
\begin{align}\label{eq2}
		\int_A^B d\Sigma.
\end{align}
The mathematical resemblance of the two previous ideas in the application of (\ref{eq1}) and (\ref{eq2}) suggests one to make the identification
\begin{align}
\label{effec}
	N_g^2(\beta)dl^2=\sum_{i,j}{g_{ij}dx^idx^j}=d\Sigma^2,
\end{align}
{where $dl^2$ is recognized} as the Euclidean line element. Still according to Joets and Ribotta, the resulting space is a Finslerian one. However, due to the symmetry of the liquid crystal molecules and the spatial configuration of the director $\hat{n}$ in the applications where this metric approach will be applied, the Riemannian geometry is enough to describe the sound propagation \cite{satiro1,moraes}.

Thus, substituting (\ref{effec}) in
(\ref{time}), we found the general expression of the line element experienced by the sound wave {in cylindrical coordinates
\begin{align}\label{analogline}
\begin{split}
ds^2=&-c^2dt^2\\&+N^2_g\left(\beta\right)\left(dr^2+r^2d\theta^2+dz^2\right)
\end{split}
\end{align}
and the covariant components of the acoustic metric related to the line element (\ref{analogline}) are 
\begin{align}\label{ino}
 \textbf{g}=\left(
\begin{array}{cccc}
 -c^2&0&0&0\\
0&N^2_g(\beta)&0&0\\
0&0&r^2N^2_g(\beta)&0\\
0&0&0&N^2_g(\beta)
\end{array}\right).
\end{align}}

{Form the standpoint of sound, an anisotropic medium can thus be mimicked by an effective gravitational field.} As for light, the WKB (or short wavelength) approximation enables to identify sound paths to the null geodesics of the effective metric
$\textbf{g}$. In pseudo-Riemannian geometry \cite{weinberg,misner,schutz}, the curve that minimizes the line element $ds^2$ is obtained from the geodesic equation
\begin{align}\label{geo}
\frac{d^2x^\mu}{d\tau^2}+{\Gamma_{\nu\sigma}^\mu\frac{dx^\mu}{d\tau}\frac{dx^\sigma}{d\tau}
} =0 ,
\end{align}
with $\tau$ a parameter along the geodesic and $\Gamma_{\nu\sigma}^\mu$ the components
of the Riemannian connection
\begin{align}\label{rc}
\Gamma_{\nu\sigma}^\mu=\frac{1}{2}g^{\mu\xi}\left(\frac{\partial g_{\xi\sigma}}{\partial x^\nu}+\frac{\partial
g_{\nu\xi}}{\partial x^\sigma}-\frac{\partial g_{\nu\sigma}}{\partial x^\xi}\right).
\end{align}

In the next section, we will apply these equations {to study sound propagation near defects occurring in the nematic phase} of liquid crystals.

\section{Applications}
\subsection{Sound paths near a defect in nematics\label{app1}}

\begin{figure}[tb]
	\centering
		\includegraphics[height=1.6in]{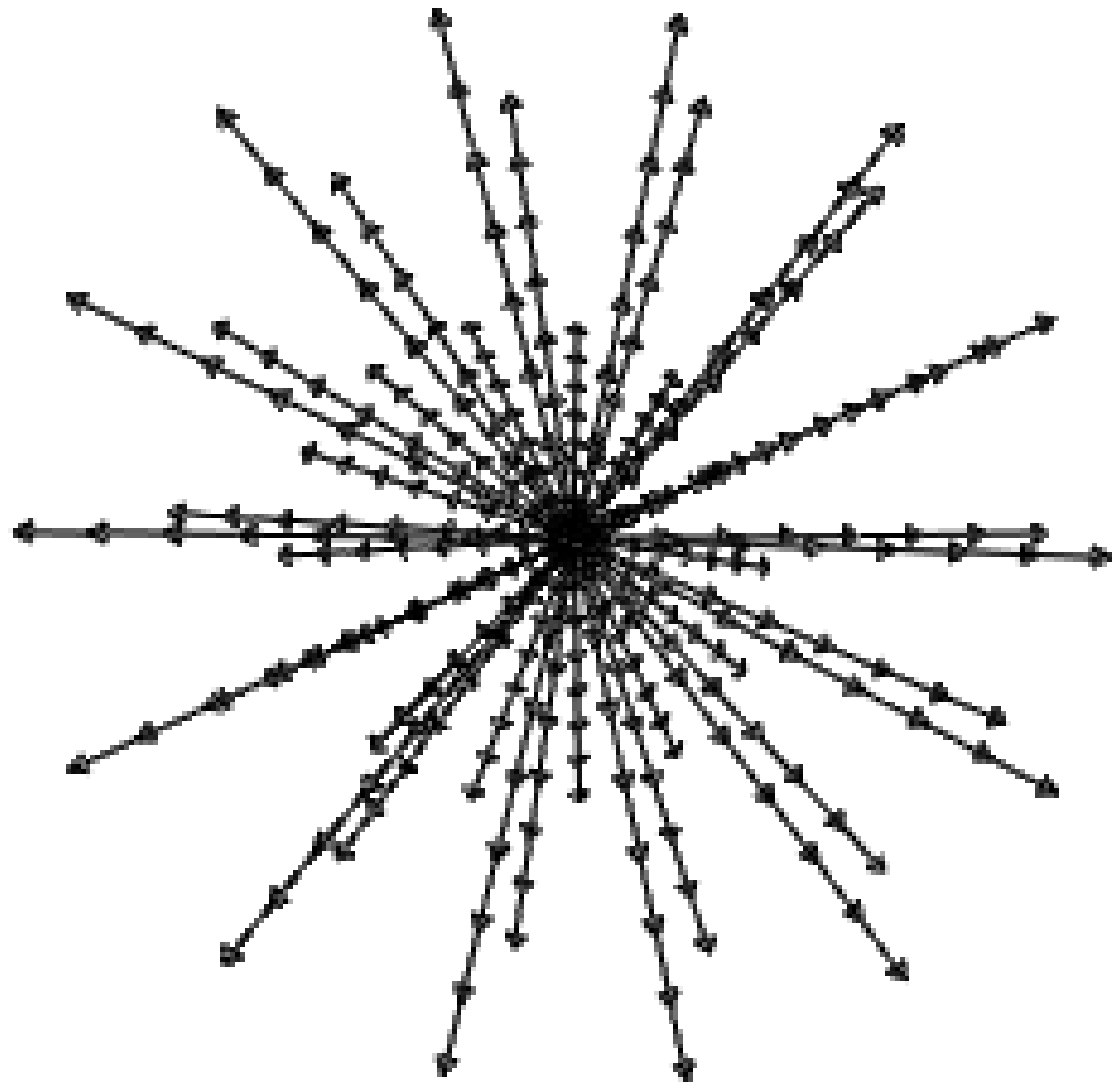}
			\includegraphics[height=1.6in]{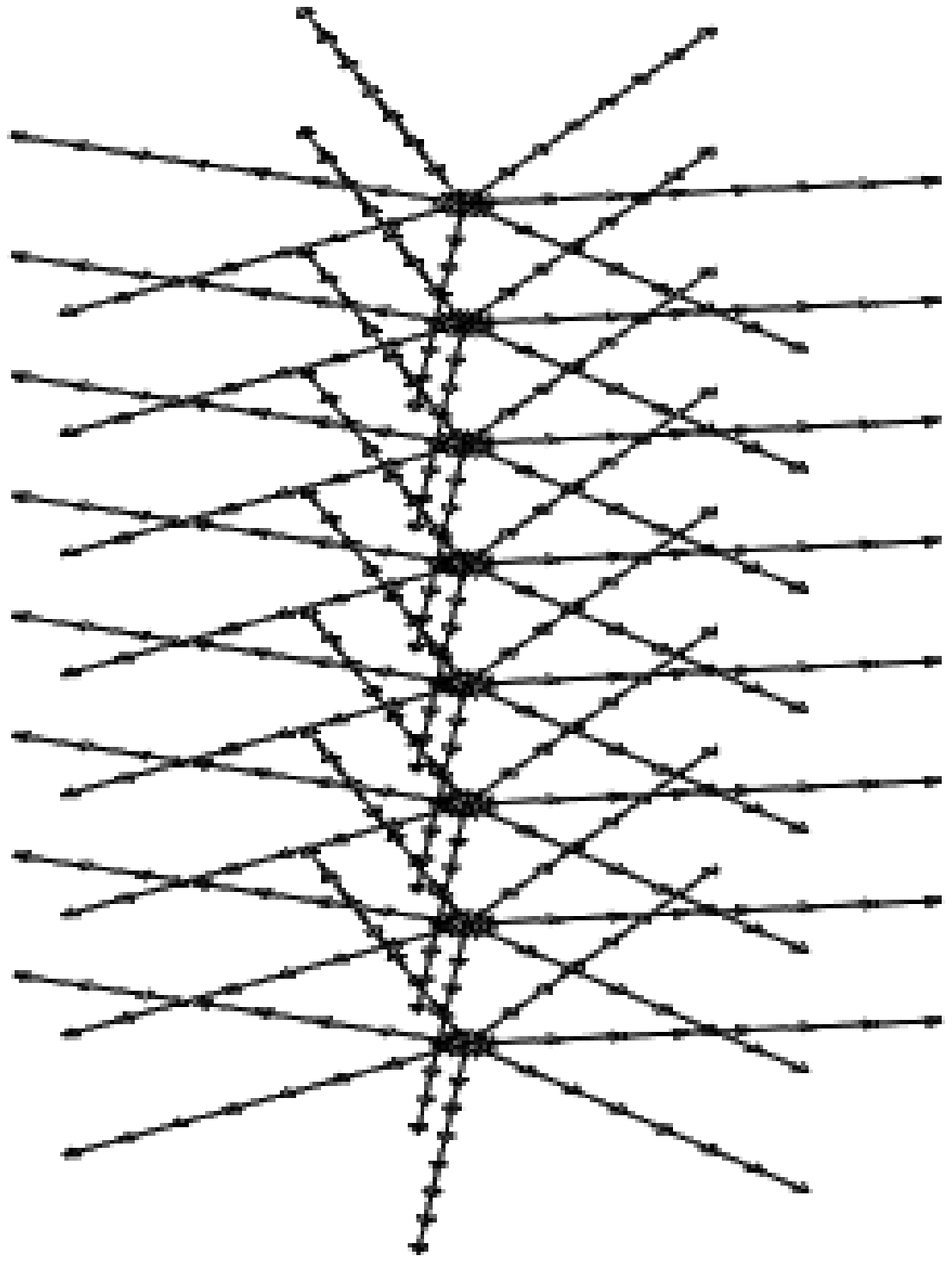}
			\caption{Hedgehog and $(k=1,c=0)$-disclination defects. The arrows represent the director $\hat{n}$.}
	\label{defects}
\end{figure}

When a liquid crystal transits from the liquid (or
isotropic) phase to nematic one, defects (points or lines)
can arise spontaneously, causing a reorientation of the director. We will focus on two different topological defects of the nematic phase: the{punctual} defect called \emph{hedgehog}, with director $\hat{n}=\hat{r}$ written in spherical coordinates, and the linear defect called \emph{disclination} with director $\hat{n}=\hat{\rho}$ written in cylindrical coordinates (Fig. \ref{defects}). 

To determine the acoustic metric associated to an hedgehog defect, the strategy is to express $\cos\beta$ and $\sin\beta$ in terms of the laboratory frame coordinates, by the Euclidean line element $dl^2=dr^2+r^2\left(d\theta^2+\sin^2\theta d\phi^2\right)$, substitute them in (\ref{extra}) and we obtain the $g_{ij}$'s by (\ref{effec}).

\begin{figure}

			\includegraphics[height=1.6in]{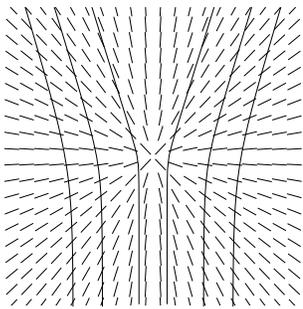}
			\caption{Sound trajectories on the equatorial plane of a hedgehog in a liquid crystal with $b=0.9$. This result is similar to sound trajectories on $z=C^{st}$ plane near to a $(k=1,c=0)$-disclination lying on the z direction with $b=0.9$.}
	\label{trajectory}
\end{figure}

In Frenet-Serret frame \cite{kamien}, the position vector along the sound path $\vec{R}=r\hat{r}$ and the tangent vector $\vec{T}(l)$  are related by
\begin{align}	\vec{T}(l)=\frac{d\vec{R}}{dl}=\frac{d\left(r\hat{r}\right)}{dl}=\frac{dr}{dl}\hat{r}+r\frac{d\hat{r}}{dl}.
\end{align}
As $\vec{T}(l)$ has the same direction as $\vec{v_g}$, then
\begin{align}
\label{cos}
	\vec{T}\cdot\hat{n}=\cos\beta=\frac{dr}{dl}\equiv\dot{r}.
\end{align}
From the modified Euclidean line element
\begin{align*}
	1=\dot{r}^2+r^2\left(\dot{\theta}^2+\sin^2\theta\dot{\phi}^2\right),
\end{align*}
one identifies
\begin{align}
\label{sin}
\sin\beta=\sqrt{r^2\left(\dot{\theta}^2+\sin^2\theta\dot{\phi}^2\right)}.
\end{align}
The replacement of (\ref{cos}) and (\ref{sin}) in (\ref{extra}) results in the line element
\begin{align*}
d\Sigma^2=\frac{\rho v^2}{C_{33}}dr^2+\frac{\rho v^2}{C_{11}}r^2\left(d\theta^2+\sin^2\theta d\phi^2\right),
\end{align*}
which gives the effective metric for sound in the vicinity of a hedgehog defect. The rescaling of the radial coordinate of this equation by $\tilde{r}\equiv\sqrt{\frac{\rho}{C_{33}}}vr$ produces
\begin{align}
\label{cs}
	d\Sigma^2=d\tilde{r}^2+b^2\tilde{r}^2\left(d\theta^2+\sin^2\theta d\phi^2\right),
\end{align}
where $b^2\equiv\frac{C_{33}}{C_{11}}$. Therefore sound paths are the geodesics of (\ref{cs}), that is the spatial part of line element of a global monopole \cite{barriola}, and they are depicted on Fig. (\ref{trajectory}) for $b=0.9$.


We now study the case of disclinations. These defects are expressed in terms of parameters $k$ and $c$ \cite{kleman}, which are related to director in cylindrical coordinates by
\begin{align}
\vec{n}=\left(\cos\left(k\phi+c\right),\sin\left(k\phi+c\right),0\right).
\end{align}
Repeating the same procedure as for hedgehog defects, we obtain a generalized effective metric, similar to one found in \cite{moraes}
\begin{eqnarray}\label{general-metric}
\nonumber d\Sigma^2&=&\left(\frac{\rho^2}{C_{33}}\cos^2\alpha+\frac{\rho^2}{C_{11}}\sin^2\alpha\right)dr^2\\
&+&\left(\frac{\rho^2}{C_{33}}\sin^2\alpha+\frac{\rho^2}{C_{11}}\cos^2\alpha\right)r^2d\phi^2\\
\nonumber &-&\left[2\left(\frac{\rho^2}{C_{11}}-\frac{\rho^2}{C_{33}}\right)\sin\alpha\cos\alpha\right]rdrd\phi,
\end{eqnarray}
where $\alpha=k\phi+c$. {In the case of ($k=1,c=0$)-disclination, the geodesics are represented on Fig. \ref{trajectory}.

Equations (\ref{cs}) and (\ref{general-metric}) are examples of effective metrics experienced by the sound close to topological defects in nematic liquid crystals. In the next section, they will serve as basis for calculations beyond the WKB approximation to determine scattering sections.

\subsection{Acoustical diffraction by defects in nematics}\label{app2}

The connection between geometric and wave optics is well-known: in the WKB approximation, light rays identify with curves that are tangent at each point to the Poynting vector \cite{born}. Similarly, since the direction of the energy velocity for sound waves defines the direction of the sound rays \cite{royer}, we use the partial wave method \cite{cohen} to examine the scattering of sound waves by the previous hedgehog defect and indicate its diffraction pattern (for disclinations, a similar analysis can be found in \cite{erms1}). Sound plane waves obey d'Alembert's wave equation for scalar fields:
\begin{equation}
\nabla_{\mu}\nabla^{\mu}\Phi\equiv\frac{1}{\sqrt{-g}}\partial_\mu\left( \sqrt{-g}g^{\mu\nu}\partial_\nu\Phi\right)=0\label{alembert}. 
\end{equation}
Here, we denote by $\partial_\mu\equiv\frac{\partial}{\partial x^\mu}$, $g_{\mu\nu}$ are the components of the metric $\textbf{g}$ (determinant $g$) given by eq. (\ref{cs}). We have also used the convention of the repeated indexes for summations. Solutions are the usual harmonic plane waves of the form $\Phi=\Phi(t,r,\theta,\phi)=e^{-i\omega t}\psi(r,\theta,\phi)$, where $\omega$ is the wave pulsation.

The spherical symmetry of the problem allow us to write $\psi(r,\theta,\phi)=\psi(r,\theta)=\sum_{l=0}^\infty{a_lR_l(r)P_l(\cos\theta)}$, where $P_l(\cos\theta)$ are the Legendre polynomials of order $l$ and $a_l$ are worthless constants for the partial wave method. Applying $\psi(r,\theta)$ in $\Phi(t,r,\theta,\phi)$ and expanding (\ref{alembert}), we obtain the radial equation 
\begin{eqnarray}\label{bessel}	
R^{''}_l(r)+2\frac{R^{'}_l(r)}{r}+\left[\omega^2-\frac{l(l+1)}{b^2r^2}\right]R_l(r)=0,
\end{eqnarray}
where $R^{'}_l(r)\equiv\partial_rR_l(r)$.
Solutions to this equation are Bessel functions of the first kind $R_l(r)=J_{n(l)}(r)$, where
\begin{eqnarray*}
	n(l)=\frac{1}{b}\left[ \left( l+\frac{1}{2}\right)^2 - \frac{1-b^2}{4}\right]^{\frac{1}{2}}.
\end{eqnarray*}
When $b=1$, we recover the flat space and the last equation becomes $l+\frac{1}{2}$. The \emph{phase shift} $\delta_l(b)$ of the scattered wave is
\begin{eqnarray}
\label{shift}
	\delta_l(b)&=&\frac{\pi}{2}\left( l+\frac{1}{2}-n(l)\right)\nonumber\\
	&=&\frac{\pi}{2}\left( l+\frac{1}{2}-\frac{1}{b}\left[ \left( l+\frac{1}{2}\right)^2 - \frac{1-b^2}{4}\right]^{\frac{1}{2}}\right).             
\end{eqnarray}

The angular distribution of the scattered sound is given by \cite{cohen} the differential scattering cross section $\sigma(\theta)$
\begin{align}\label{dif}
\sigma(\theta)\equiv\left|f(\theta)\right|^2=\left|\frac{1}{2i\omega}\sum_{l}^{\infty}(2l+1)(e^{2i\delta_{l}}-1)P_{l}(\cos\theta)\right|^2,
\end{align}
where $f(\theta)$ is called scattering amplitude. The spherical symmetry of the scattering generates an annular diffraction pattern, a ring of sound. The angular location of the maximum of sound is given by eq. (\ref{dif}), as it is shown in Fig. (\ref{fig:geral1}).
\begin{figure}
	\centering
		\includegraphics[height=2.5in]{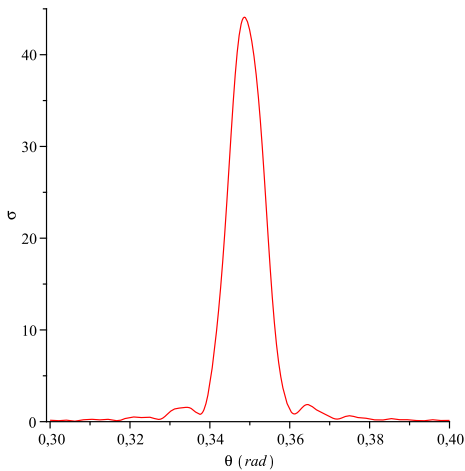}
	\caption{Differential scattering cross section, eq. (\ref{dif}), for a hedgehog defect with $\left(b=0.9,\omega=1\right)$ truncated at $l=600$.}
	\label{fig:geral1}
\end{figure}

An analytical expression for the location of the light diffraction ring scattered by a global monopole defect in the real spacetime is found in \cite{mazur} and a similar one is developed here. The idea is to expand the scattering amplitude about $b^2\approx1$ 
\begin{align}
	f(\theta)=f^{(0)}(\theta)+f^{(1)}(\theta)+\ldots,
\end{align}
and to analyze the angular behavior of the first two terms. For the phase shift (\ref{shift}), we make $\zeta \equiv l+ \frac {1} {2} $ and $a^2 \equiv \frac {1-b^2} {4} $, obtaining
\begin{equation*}
\delta_l(b)=\frac{\pi}{2}\left(\zeta-\frac{\zeta}{b}\sqrt{1-\frac{a^2}{\zeta^2}}\right).
\end{equation*}
We expand this last equation about $a^2\approx0$ (that is equivalent to expand about $b^2\approx1$), resulting in
\begin{eqnarray}
\label{fase}
 \delta_l(b)\approx\frac{\pi}{2}\left[\left(1-\frac{1}{b}\right)\zeta+\frac{a^2}{2b\zeta}+O(a^4)
 \right].
\end{eqnarray}
Substituting (\ref{fase}) in the scattering amplitude, we obtain the first two terms of the expansion, $f^{(0)}(\theta)$ and $f^{(1)}(\theta)$. They can be written in terms of the generating functions, $h(\theta,\alpha)$, of the Legendre polynomials
\begin{eqnarray*}
	h(\theta,\alpha)=\sum_{l=0}^{\infty}{e^{\pi i\alpha \zeta(l)}P_l(\cos\theta)} =\frac{1}{\sqrt{2(\cos\pi\alpha-\cos\theta)}}, 
\end{eqnarray*}
where $\alpha=1-\frac{1}{b}$. The zero-order term, $f^{(0)}(\theta)$, can be written considering the derivative of $h(\theta,\alpha)$
\begin{eqnarray}
\label{f0}	f^{(0)}(\theta)=\frac{1}{2\sqrt{2}\omega}\frac{\sin\pi\alpha}{(\cos\pi\alpha-\cos\theta)^{3/2}}, 
\end{eqnarray}
and the first-order term, $f^{(1)}(\theta)$, can be written considering $h(\theta,\alpha)$
\begin{eqnarray}
\label{f1}			  	
f^{(1)}(\theta)=\frac{\pi\alpha^2}{2b\omega}\frac{1}{\sqrt{2(\cos\pi\alpha-\cos\theta)}}.
\end{eqnarray}

The equations (\ref{f0}) and (\ref{f1}) diverge when $\theta=\theta_0=\pi\alpha$, that is the analytical expression for the angular location of the diffraction ring. For example, when $b=0.9$, $\theta_0=\pi\left(1-1/0.9\right)\approx0.35$ rad, as indicated by Fig. (\ref{fig:geral1}). For $(k=1,c=0)$ disclinations, the same result is obtained \cite{erms1} and they are in  agreement with the calculations derived by Grandjean \cite{grandjean}.

\section{Conclusion and Perspectives}
We have developed an effective geometry approach to investigate properties of sound propagation inside anisotropic media such as nematic liquid crystals. In this framework, the influence of each defect is represented by an effective metric: metrics for hedgehogs are similar to those of global monopoles, whereas (k, c)-disclinations have more complex forms. Sound trajectories and diffraction patterns are thus modified by the non-trivial metric. In particular, sound scattering by hedgehog defects is identical to light scattering by a global monopole. 

It's interesting to note the relation between the metric approach shown here and Katanaev-Volovich's theory \cite{katanaev1}. While Katanaev uses an affine transformation to deform and curve the elastic medium, generating the anisotropic properties, we start from the anisotropic velocity to derive the effective metric. 

We could reinterpretate the obtained results by Cosmology's view. Once sound waves can be understood as propagating perturbations in the effective metric generated by the director field  $\hat{n}(\vec{r})$, they are analog to gravitational waves propagating in a true gravitational metric generated by a cosmic defect . Thus, the annular diffraction pattern due to defects in a liquid crystal can also be expected for gravitational waves coming on cosmic strings or global monopoles. In other words, the diffraction pattern is a gravitational signature of the presence of cosmic defects, and its properties may help retrieving information about them.  Furthermore, considering the weak effects of gravitational waves on possible detectors, we have presented an algebraic result for the angle of the maximum intensity of the scattered gravitational waves. Since the scattered wave on this angle has amplitude greater than incident wave's one, our results indicate the angular position where detectors of gravitational waves must be place to improve their power of detection. Another possible extension in this way is the possibility to test a cosmological mechanisms predicted in the paper of T. Damour and A. Vilenkin \cite{damour}: from our analogy, production of sound waves should occur by cusps of disclinations in liquid crystals. If required by the referee, we can insert these comments in our article.

In the language of cohomology \cite{cohomology}, the Volterra process for a cosmic string is obtained by a removing of a cylindrical solid angle, whereas the Volterra process of the corresponding antistring is obtained by an addition of a cylindrical solid angle. For example, in nematics, such pair will consist in a $(k=1,c=0)$-disclination, with $b^2<1$ in its effective metric, being the analogue for a cosmic string, and a $(k=1,c=\pi/2)$-disclination, with $b^2>1$ in its effective metric, being the analogue for the anti cosmic string. As it is known, the annihilation of disclinations in nematics, a velocity field is expected near the pairs, keeping the director fixed \cite{svensek,toth,dierking}. And, as in the analogue models based in moving fluid where the gradient of velocity field maps the gravitational field \cite{novello}, we could previse the presence of a gravitational field in the vicinity of two annihilating cosmic strings. 

Still in cosmology, topological defects such as cosmic strings and global monopoles appear as a result of symmetry breakings during phase transitions occurring while the early universe cooled down. In the Kibble mechanism \cite{kibble}, the formation of these defects gives rise to emission of gravitational waves. In liquid crystals, defects appear as a result of SO(3) symmetry breaking of the isotropic-nematic transition phase during cooling. Formation of defects also obeys a Kibble mechanism \cite{bowick,breno}. Therefore, by means of our analogy, we should be able (in principle) to characterize the properties of gravitational waves (such as polarization for example) from the study of sound waves propagating in nematic liquid crystals.

Future developments of this work concern on the influence of temperature on elastic constants, directors fluctuations (corresponding to fluctuations of the metric itself \cite{svaiter}) and consequently on the sound trajectories and diffractions. Another interesting path should be the possibility of simulating the exterior metric of spherical, uncharged, non-rotating mass, known as Schwarzschild metric \cite{misner,weinberg}, for sound rays (as it was proposed for light in \cite{erms2}).

\bibliography{pereira}

\end{document}